\documentclass[letterpaper, 10 pt, conference]{ieeeconf}  

\IEEEoverridecommandlockouts                              

\usepackage{graphicx}
\usepackage{amsmath} 
\usepackage{amssymb}  
\usepackage{booktabs}
\usepackage{threeparttable}
\usepackage{cite}

\newcommand{\argmax}{\mathop{\rm arg\, max}\limits}

\title{\LARGE \bf
A Time-Series Scale Mixture Model of EEG with a Hidden Markov Structure for Epileptic Seizure Detection
}

\author{Akira Furui$^{1}$, \textit{Member, IEEE}, Tomoyuki Akiyama$^{2}$, and Toshio Tsuji$^{1}$, \textit{Member, IEEE}
\thanks{This work was partially supported by Grant-in-Aid for JSPS Research Fellows 18J22370.}
\thanks{$^{1}$A. Furui and T. Tsuji are with the Graduate School of Advanced Science and Engineering, Hiroshima University, Higashihiroshima 739-8527, Japan (e-mail: akirafurui@hirosima-u.ac.jp).}
\thanks{$^{2}$T. Akiyama is with the Department of Child Neurology, Okayama University Hospital, Okayama 700-8558, Japan.}}%

\begin{document}

\maketitle
\thispagestyle{empty}
\pagestyle{empty}

\begin{abstract} 
In this paper, we propose a time-series stochastic model based on a scale mixture distribution with Markov transitions to detect epileptic seizures in electroencephalography (EEG). In the proposed model, an EEG signal at each time point is assumed to be a random variable following a Gaussian distribution. The covariance matrix of the Gaussian distribution is weighted with a latent scale parameter, which is also a random variable, resulting in the stochastic fluctuations of covariances. By introducing a latent state variable with a Markov chain in the background of this stochastic relationship, time-series changes in the distribution of latent scale parameters can be represented according to the state of epileptic seizures. In an experiment, we evaluated the performance of the proposed model for seizure detection using EEGs with multiple frequency bands decomposed from a clinical dataset. The results demonstrated that the proposed model can detect seizures with high sensitivity and outperformed several baselines.
\end{abstract}

\vspace{-2mm}

\section{Introduction}
Epilepsy is a heterogeneous neurological disorder characterized by a transient abnormal discharge of neurons.
This disorder involves recurrent and unprovoked seizures called epileptic seizures~\cite{Kandel2013-xu}.
The most common method for detecting epileptic seizures is electroencephalography (EEG) recorded from the scalp.
However, the detection of seizures in clinical practice relies on visual inspection, which requires a high level of expertise, and also places a heavy burden on the epileptologist because of the need for prolonged observation.

Various attempts have been made to automatically detect epileptic seizures from EEG~\cite{Shoeb2010-hp,Greene2008-yz,Subasi2007-bz,Wong2007-ax,Craley2021-mn}.
Many of these studies focused on capturing the features that reflect the specific morphologies of seizure activity from EEG.
The effectiveness of time-domain methods, such as the root mean square (RMS) and entropy~\cite{Greene2008-yz}, and time-frequency domain methods, such as wavelet-based features~\cite{Subasi2007-bz}, has been demonstrated.

There is a time-dependent evolution of physiological states behind the changes in EEG characteristics with the onset and progression of seizures.
To capture such a non-stationarity of EEG, its dynamic characteristics should be directly considered.
Several previous studies have used machine learning techniques that take into account the time-series nature of EEG for seizure detection, and have outperformed framewise static classifiers~\cite{Wong2007-ax,Craley2021-mn}.

Meanwhile, we previously focused on the non-Gaussianity of EEG and proposed a stochastic model based on the scale mixture distribution~\cite{Furui2021-rc} called the scale mixture model (SMM).
This model can represent the stochastic fluctuations latent in the amplitude of EEG.
We have shown that an index defined based on the model may be more effective than conventional time-domain features in detecting epileptic seizures.
However, this model is static in nature and cannot track the dynamic properties of time-series EEG.

This paper proposes a discrete state-space EEG model that incorporates the scale mixture distribution and apply it to epileptic seizure detection.
The proposed model, referred to as the hidden Markov SMM (HMSMM), is capable of representing the distribution changes between the non-seizure and seizure states of EEG by switching the distribution of the latent scale parameter based on Markov transitions.
We evaluated the detection performance of the proposed HMSMM experimentally using a clinical EEG dataset.

\section{Hidden Markov Scale Mixture Model (HMSMM) of EEG}

\subsection{Model structure}

\begin{figure}[!t]
   \centering
   \includegraphics[width=0.7\hsize]{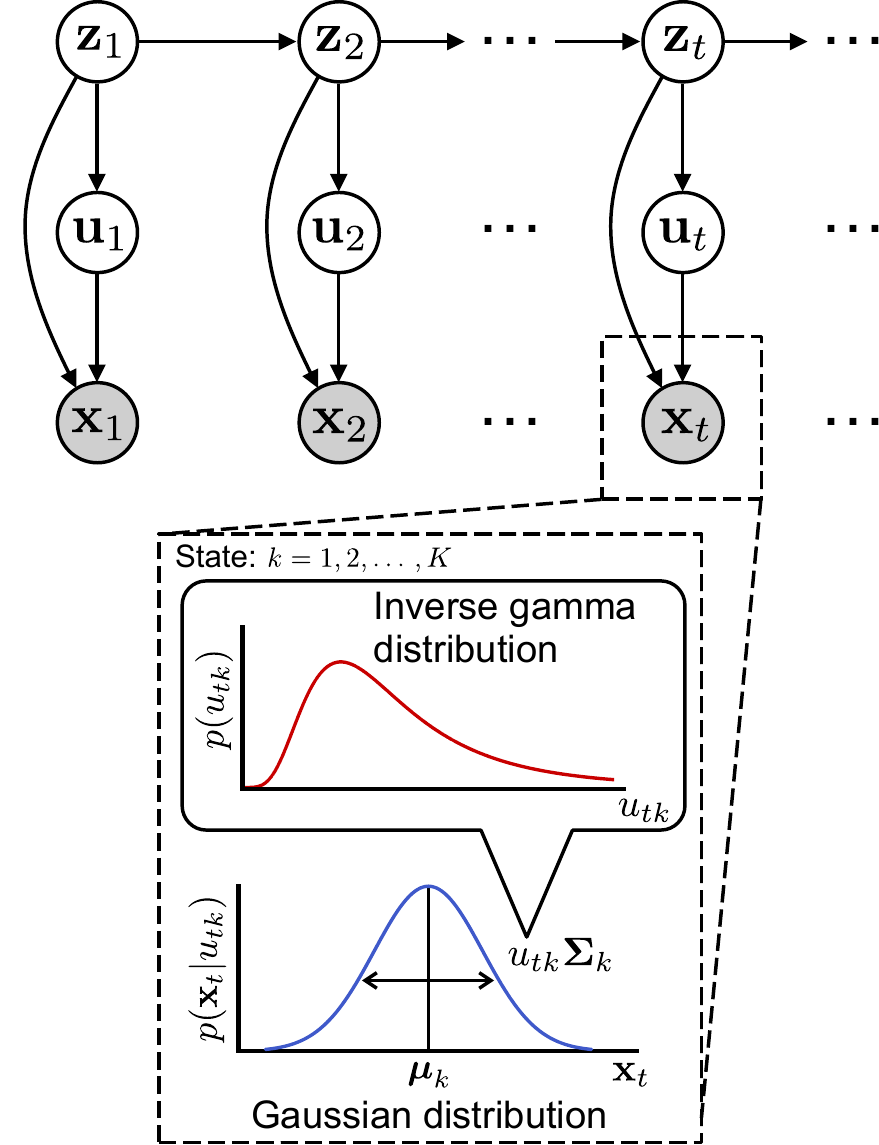}
   \caption{Graphical model of the proposed hidden Markov scale mixture model (HMSMM).
   The shaded and white circles represent the observed and latent random variables, respectively.}
   \label{fig:model}
\end{figure}

Fig.~\ref{fig:model} shows a graphical model of the proposed HMSMM.
The state transitions existing in the background of EEG generation are represented by a discrete latent variable $\mathbf{z}_t$.
The emission distribution from each state is based on the SMM.
In this model, EEG $\mathbf{x}_t \in \mathbb{R}^D$ ($D$ is the number of electrodes) at time $t$ from each state is represented by a conditional Gaussian distribution.
The covariance matrix of the Gaussian distribution is weighted by a latent scale parameter $u_{tk}$ ($k = 1,2,\ldots,K$; $K$ is the number of states), which is also a random variable.
The stochastic behavior of $u_{tk}$ switches with state transitions, and its randomness causes the covariance matrix to fluctuate at each time point, which results in the modulation of the Gaussianity of the EEG.

For state $k$, the EEG signal $\mathbf{x}_t$ is assumed to be generated from the following scale mixture distribution:
\begin{align}
   p(\mathbf{x}_t|&{z}_{tk} = 1) \nonumber \\
   & = \int \mathcal{N}(\mathbf{x}_t|\boldsymbol{\mu}_k, u_{tk}\mathbf{\Sigma}_k) \mathrm{IG}(u_{tk}|\nu_k/2, \nu_k/2) \mathrm{d}u_{tk},
\end{align}
where $\mathbf{z}_t = \{z_{tk} \}$ is the latent variable based on the 1-of-$K$ representation, and assigns observations to each state.
$\mathcal{N}(\mathbf{x}_t|\boldsymbol{\mu}_k, u_{tk}\mathbf{\Sigma}_k)$ denotes the Gaussian distribution:
\begin{align}
   \mathcal{N}(\mathbf{x}_t|&\boldsymbol{\mu}_k, u_{tk}\mathbf{\Sigma}_k) \nonumber \\
      &= (2\pi)^{-\frac{D}{2}} |u_{tk} \mathbf{\Sigma}_k|^{-\frac{1}{2}} \exp \left[-\frac{u_{tk}^{-1}}{2} d(\mathbf{x}_t; \boldsymbol{\mu}_k, \mathbf{\Sigma}_k) \right],
\end{align}
where $\boldsymbol{\mu}_k$ and $\mathbf{\Sigma}_k$ are the mean vector and covariance matrix, respectively, $d(\mathbf{x}_t; \boldsymbol{\mu}_k, \mathbf{\Sigma})$ is the squared Mahalanobis distance, and $u_{tk}$ is the latent scale parameter following an inverse gamma distribution given by
\begin{align}
   \mathrm{IG}(u_{tk}|a_k, b_k) = \frac{b_k^{a_k}}{\Gamma(a_k)} (u_{tk})^{-a_k-1} \exp \left[-\frac{b_k}{u_{tk}} \right].
\end{align}

The state transitions from $\mathbf{z}_{t-1}$ to $\mathbf{z}_{t}$ can be represented by the conditional distribution $p(\mathbf{z}_{t}|\mathbf{z}_{t-1})$.
Because $\mathbf{z}_{t}$ is $K$-dimensional binary variable, this conditional distribution is denoted by the matrix form $\mathbf{A} \in \mathbb{R}^{K \times K}$, which elements are transition probabilities given by
\begin{align}
   p(\mathbf{z}_t|\mathbf{z}_{t-1}) = \prod_{k=1}^K \prod_{j=1}^J A_{jk}^{z_{t-1,j}{z_{tk}}},
\end{align}
where $0 \leq A_{jk} \leq 1$ and $\sum_k A_{jk} = 1$.
The initial state $\mathbf{z}_1$ is a special case because the previous state does not exist; therefore $\mathbf{z}_1$ is expressed as
\begin{align}
   p(\mathbf{z}_1) = \prod_{k=1}^K \pi_k^{z_{1k}},
\end{align}
where $\sum_{k} \pi_k = 1$.

\subsection{Inference of states}
Our goal is to predict the latent seizure states by calculating the posterior distribution $p(\mathbf{z}_t|\mathbf{X})$ given an observed EEG sequence $\mathbf{X} = \{\mathbf{x}_1, \ldots, \mathbf{x}_T \}$.
We can calculate this efficiently based on the forward-backward algorithm~\cite{Bishop2006-th} as follows:
\begin{align}
   p(\mathbf{z}_t|\mathbf{X}) &= \frac{p(\mathbf{X}|\mathbf{z}_t) p(\mathbf{z}_t)}{p(\mathbf{X})} \nonumber \\
   &= \frac{p(\mathbf{x}_1, \ldots, \mathbf{x}_t, \mathbf{z}_t) p(\mathbf{x}_{t+1}, \ldots, \mathbf{x}_T|\mathbf{z}_t)}{p(\mathbf{X})} \nonumber \\
                              &= \frac{\alpha(\mathbf{z}_t) \beta(\mathbf{z}_t)}{p(\mathbf{X})},
\label{eq:pred}
\end{align}
where $\alpha(\mathbf{z}_t)$ is calculated by the forward recursive calculation given by
\begin{align}
   \alpha(\mathbf{z}_t) &= p(\mathbf{x}_t|\mathbf{z}_t) \sum_{\mathbf{z}_{t-1}} \alpha(\mathbf{z}_{t-1}) p(\mathbf{z}_t|\mathbf{z}_{t-1}), \\
   \alpha(\mathbf{z}_1) &= p(\mathbf{z}_1)p(\mathbf{x}_1|\mathbf{z}_1) = \prod_{k=1}^K \{ \pi_k p(\mathbf{x}_1)\}^{z_{1k}},
\end{align}
and $\beta(\mathbf{z}_t)$ is calculated by the backward recursive calculation as follows:
\begin{align}
   \beta(\mathbf{z}_t) &= \sum_{\mathbf{z}_{t+1}} \beta(\mathbf{z}_{t+1}) p(\mathbf{x}_{t+1}|\mathbf{z}_{t+1}) p (\mathbf{z}_{t+1}|\mathbf{z}_t).
\end{align}
The denominator of (\ref{eq:pred}) is given by $p(\mathbf{X}) = \sum_{\mathbf{z}_T} \alpha(\mathbf{z}_T)$.
To make predictions for novel data, the parameters of the state transition and emission distribution must be learned from the given training data in advance.
In the next subsection, we outline the learning algorithm of the proposed model.

\subsection{Learning algorithm}
Let a set of EEG sequences $\mathcal{D} = \{\mathbf{X}_1, \ldots, \mathbf{X}_N\}$, where $\mathbf{X}_n = \{\mathbf{x}_{nt}\}_{t=1}^{T_n}$ is an $n$-th sequence of length $T_n$, be the training set.
Assuming that the hidden state sequences $\{\mathbf{z}_{nt}\}$ corresponding to the EEG sequences $\mathcal{D}$ are available at the training stage (e.g., clinical annotation of seizures by experts), the model can be trained in a supervised manner.

\subsubsection{Update the transition parameters}
In this step, the transition probabilities are calculated between states.
The initial and transition probabilities can be updated as follows:
\begin{align}
   \pi_k = \frac{N^1_k}{\sum_{k=1}^{K} N^1_k}, \quad
   A_{jk} = \frac{N_{jk}}{\sum_{k=1}^K N_{jk}},
\end{align}
where $N^1_k$ is the number of times that the initial state is $k$ in all sequences, and $N_{jk}$ is the number of transitions from state $j$ to $k$ in all sequences.

\subsubsection{Update the emission distribution}
The parameters of the emission distribution for each state are estimated based on the expectation-maximization (EM) algorithm~\cite{Bishop2006-th,Furui2021-rc}.
The E-step calculates the expectation of the complete-data log-likelihood:
\begin{align}
   Q(\nu_k, \boldsymbol{\mu}_k, \mathbf{\Sigma}_k) = \mathbb{E} \biggl[\ln &\prod_{n=1}^{N} \prod_{t=1}^{T_n} \prod_{k=1}^{K} z_{ntk}\, \mathcal{N}(\mathbf{x}_{nt}|\boldsymbol{\mu}_k, u_{ntk} \mathbf{\Sigma}_k) \nonumber \\
   \quad &\times \mathrm{IG}(u_{ntk}|\nu_k/2, \nu_k/2) \biggr].
\end{align}
The following expectation is calculated using the current parameters:
\begin{align}
   \tau_{ntk} \triangleq  \mathbb{E}[u^{-1}_{ntk}] = \frac{\nu_k + D}{\nu_k + d(\mathbf{x}_{nt}, \boldsymbol{\mu}_k, \mathbf{\Sigma}_k)}.
\end{align}
In the M-step, $Q(\nu_k, \boldsymbol{\mu}_k, \mathbf{\Sigma}_k)$ is maximized with respect to each parameter, and the update equations are obtained:
\begin{align}
   ^{\mathrm{new}}\boldsymbol{\mu}_k = \frac{\sum_{n=1}^N \sum_{t=1}^{T_n} z_{ntk}\, \tau_{ntk} \mathbf{x}_{nt}}{\sum_{n=1}^N \sum_{t=1}^{T_n} z_{ntk} \tau_{ntk}},
\end{align}
\begin{align}
   ^{\mathrm{new}}\mathbf{\Sigma}_k = \frac{\sum_{n=1}^N \sum_{t=1}^{T_n} z_{ntk}\, \tau_{ntk} (\mathbf{x}_{nt} - \boldsymbol{\mu}_k)(\mathbf{x}_{nt} - \boldsymbol{\mu}_k)^{\top}}{\sum_{n=1}^N \sum_{t=1}^{T_n} z_{ntk}}.
\end{align}
The parameter $\nu_k$ is estimated by iteratively maximizing $Q(\nu_k, \boldsymbol{\mu}_k, \mathbf{\Sigma}_k)$ using the bisection method:
\begin{align}
   ^{\mathrm{new}}\nu_k = \argmax_{\nu_k} Q(\nu_k, \boldsymbol{\mu}_k, \mathbf{\Sigma}_k).
\end{align}
The EM iteration repeats until the calculation converges.

\section{Experiments}
We evaluated our model on the clinical EEG dataset used in \cite{Furui2021-rc}.
The dataset contains scalp EEG recordings from 20 epileptic patients with focal epilepsy (age: 0.5 to 41 years).
The EEG data were recorded with a sampling frequency of 500 Hz.
The 19-channel surface electrodes ($D=19$) were placed on the scalp according to the international 10--20 electrode system, with reference electrodes on both earlobes: A1 and A2.
Each patient had a single EEG sequence containing the seizure duration.
The average data length and seizure duration were $315.0 \pm 42.4$ s and $48.8 \pm 23.7$ s, respectively.
The experiments were approved by the Okayama University Ethics Committee (approval No: 1706-019).
The onset and offset of a seizure in each EEG recording were marked by a board-certified epileptologist.

EEG has different characteristics depending on the frequency band.
Therefore, the EEG signals were decomposed into five frequency bands using a filter bank: $\delta$ (1--3 Hz), $\theta$ (4--7 Hz), $\alpha$ (8--12 Hz), $\beta$ (13--24 Hz), and $\gamma$ (25--80 Hz).
These are the common frequency bands in EEG analysis.
To reduce differences in the EEG amplitude range caused by scalp impedance and other factors for each subject, each EEG sequence was normalized by dividing it by the standard deviation of its first 5 s.
For post-processing, the estimated latent state sequence was smoothed by averaging them with a moving window of length 5 s.

In this experiment, three states ($K=3$) were considered: pre-seizure (state 1), seizure (state 2), and post-seizure (state 3).
The transition probabilities between these latent states were constrained as follows:
\begin{equation}
   A_{jk} = 
      \left[
         \begin{array}{rrr}
            a_{11}   & a_{12}    & 0 \\
            0        & a_{22}    & a_{23} \\
            a_{31}   & 0         & a_{33}
            
         \end{array}
      \right].
\label{eq:our_A}
\end{equation}
This matrix structure ensured that state transitions were only allowed in the order of pre-seizure, seizure, and post-seizure, and after post-seizure, the state returned to pre-seizure.
Only the non-zero elements in (\ref{eq:our_A}) were estimated through training.

The performance of the proposed model was compared with that of the Gaussian-based hidden Markov model (GHMM), original SMM~\cite{Furui2021-rc}, linear logistic regression (LLR), and multi-layer perceptron (MLP).
The same pre- and post-processing as that in the proposed model were performed for these baseline models.
The GHMM has the same time-series structure as the proposed model, but the emission distribution is replaced by a Gaussian distribution.
The SMM corresponds to the proposed model without the time-series structure.
Each learned SMM $p(\mathbf{x}|\mathbf{z}_{t})$ was constructed by fitting the model for each state, and the seizure probability (state 2) was calculated based on
\begin{align}
   p(z_{t2} = 1|\mathbf{x}_{t}) = \frac{p(z_{t2} = 1) p(\mathbf{x}_{t}|z_{t2} = 1)}{\sum_{k} p(z_{tk}) p(\mathbf{x}_{t}|z_{tk})},
\end{align}
where $p(z_{tk})$ was fixed to the proportion of each state in the dataset.
The MLP had a single hidden layer with 15 units and was trained using stochastic gradient descent with a batch size of 256 and learning rate of 0.001.
A weight decay of $1.0 \times 10^{-5}$ was used.
The discriminative models, LLR and MLP, could not be trained well for each EEG band and achieved very poor prediction performance.
Thus, to stabilize these inferences, amplitude features were extracted by calculating the RMS for each EEG band and channel, and used to evaluate the performance of LLR and MLP.
The RMS is the one of the most common features that characterizes the amplitude information of EEGs~\cite{Greene2008-yz}.
The RMS was calculated continuously using a moving window of length 2~s. 

\begin{figure*}[!t]
   \centering
   \includegraphics[width=0.9\hsize]{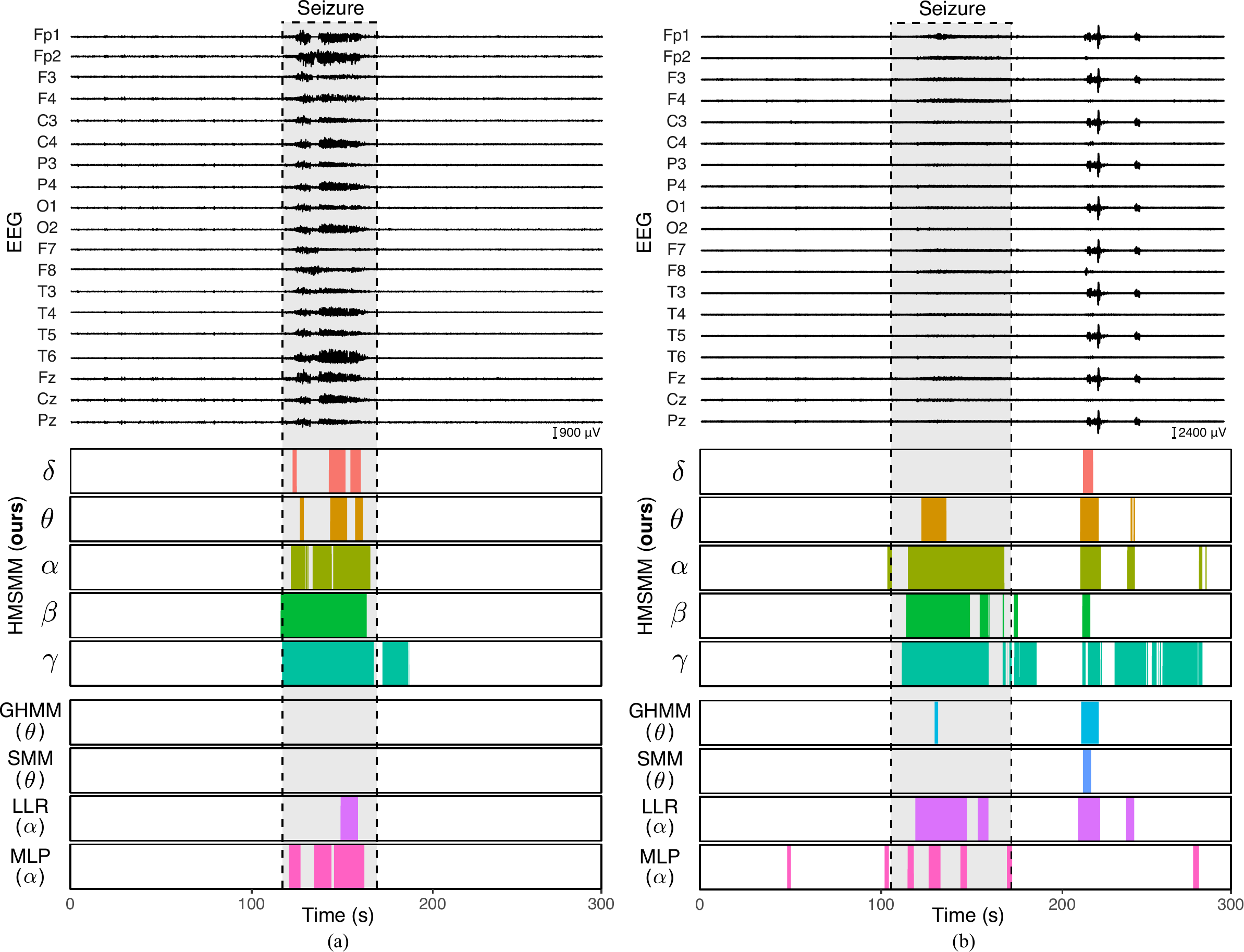}
   \caption{Raw EEG signals and corresponding detection results from the proposed HMSMM and baselines. 
   (a) Patient A. (b) Patient B.
   Seizure onset and offset are indicated by the vertical dashed lines.
   The results of the proposed HMSMM are shown for all bands.
   The results of the baseline models are shown only for the band with the highest MCC across the validation.
   }
   \label{fig:exp}
\end{figure*}

We used leave-one-patient-out cross-validation to evaluate the models. 
Data from a single patient were used as a test set and data from the remaining patients were used as the training set for the models. 
This validation procedure can be used to evaluate the generalization performance of a trained model on new patients.
For each model, the segments for which the output seizure probabilities exceeded 0.5 were defined as the detected seizure segments and the reminder as the non-seizure segments.
Seizure detection performance was quantitatively evaluated using five metrics: sensitivity, specificity, Matthews correlation coefficient (MCC), area under the receiver operating curve (AUC-ROC), and area under the precision-recall curve (AUC-PR).
Sensitivity and specificity are measures of the proportion of seizure and non-seizure that are correctly classified for all prediction results, respectively.
The MCC is essentially a correlation coefficient between the true and predicted binary classification, with a value between $-1$ and $+1$.
This metric is generally regarded as a balanced measure because it can consider all true and false positives and negatives.
The AUC-ROC and AUC-PR are measures calculated from the ROC and PR curve, respectively, and provide summary scores that capture behavior at a range of detection thresholds.
Each metric was averaged across all validation sets.

\section{Results}

\begin{table*}[!t]
   \centering
   \caption{Quantitative results for each model}
   \begin{threeparttable}
   \begin{tabular}{l@{\hspace{1.2cm}}l@{\hspace{0.75cm}}l@{\hspace{0.75cm}}l@{\hspace{0.75cm}}l@{\hspace{0.75cm}}l}
   \toprule
   Model (\textit{band})               & Sensitivity     & Specificity     & MCC        & AUC-ROC         & AUC-PR          \\ 
   \midrule
   \textbf{Baseline model}$^\dagger$ \\
   \quad GHMM ($\theta$)  & 0.276 $\pm$ 0.329         & 0.981 $\pm$ 0.029         & 0.290 $\pm$ 0.342         & 0.838 $\pm$ 0.195         & 0.613 $\pm$ 0.305         \\
   \quad SMM ($\theta$)   & 0.202 $\pm$ 0.283         & \textbf{0.984 $\pm$ 0.033} & 0.231 $\pm$ 0.297        & 0.837 $\pm$ 0.215         & 0.599 $\pm$ 0.294         \\
   \quad LLR ($\alpha$)     & 0.302 $\pm$ 0.359         & 0.983 $\pm$ 0.032         & 0.310 $\pm$ 0.367         & 0.842 $\pm$ 0.177         & 0.629 $\pm$ 0.290         \\
   \quad MLP ($\alpha$)   & 0.442 $\pm$ 0.402         & 0.892 $\pm$ 0.119         & 0.308 $\pm$ 0.321         & 0.705 $\pm$ 0.268         & 0.512 $\pm$ 0.340         \\
   \midrule 
   \textbf{Our model} \\
   \quad HMSMM ($\delta$) & 0.374 $\pm$ 0.304         & 0.905 $\pm$ 0.122         & 0.300 $\pm$ 0.284         & 0.746 $\pm$ 0.197         & 0.444 $\pm$ 0.254         \\
   \quad HMSMM ($\theta$) & 0.502 $\pm$ 0.369         & 0.910 $\pm$ 0.138         & 0.420 $\pm$ 0.331         & 0.836 $\pm$ 0.220         & 0.617 $\pm$ 0.310         \\
   \quad HMSMM ($\alpha$) & \textbf{0.631 $\pm$ 0.285} & 0.890 $\pm$ 0.143         & \textbf{0.519 $\pm$ 0.291} & \textbf{0.868 $\pm$ 0.167} & 0.654 $\pm$ 0.308         \\
   \quad HMSMM ($\beta$)  & 0.520 $\pm$ 0.396         & 0.946 $\pm$ 0.073         & 0.453 $\pm$ 0.337         & 0.864 $\pm$ 0.180         & \textbf{0.671 $\pm$ 0.296} \\
   \quad HMSMM ($\gamma$) & 0.556 $\pm$ 0.365         & 0.806 $\pm$ 0.248         & 0.347 $\pm$ 0.320         & 0.760 $\pm$ 0.264         & 0.570 $\pm$ 0.316         \\ 
   \bottomrule    \\ 
   \end{tabular}
   \vspace{-3mm}
   \begin{tablenotes}[para,flushleft]
		$^\dagger$The results of the baseline models are shown only for the band with the highest MCC across the validation.
	\end{tablenotes}
\end{threeparttable}
\label{tab:results}
\end{table*}

Fig.~\ref{fig:exp} shows an example of the seizure detection results for two patients.
The top and bottom panels present the raw EEG signals for each channel and the time-series detection results of seizures obtained by the proposed and baseline models, respectively.
The areas surrounded by black dashed lines indicate epileptic seizure occurrences diagnosed by an epileptologist.
Note that the proposed HMSMM shows the results for all frequency bands ($\delta$--$\gamma$), whereas each baseline model only shows the results for the band with the highest overall MCC.

\begin{figure}[!t]
   \centering
   \includegraphics[width=0.75\hsize]{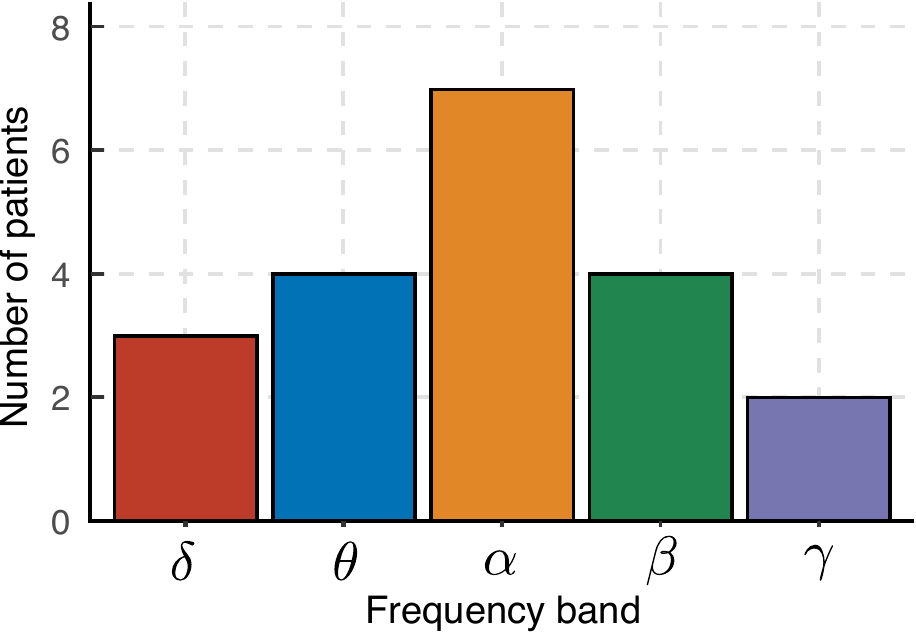}
   \caption{Number of patients with the highest MCC in each band}
   \label{fig:hist}
\end{figure}

Table~\ref{tab:results} summarizes the mean and standard deviations of the evaluation metrics for each model obtained through leave-one-patient-out cross-validation.
As in Fig.~\ref{fig:exp}, for the baseline models, only the results for the band with the highest MCC are shown.
The maximum value for each metric appears in bold.
The proposed HMSMM tended to outperform the baseline models for all metrics except specificity, particularly in the $\alpha$ and $\beta$ bands.
Fig.~\ref{fig:hist} shows the results of counting the frequency band with the highest MCC in the detection of each subject based on the cross-validation of the proposed model.
For example, the count of the $\alpha$ band was 7, which means that the highest MCC in the $\alpha$ band was obtained in 7 out of 20 patients.

\section{Discussion}

In the proposed model, the detection results corresponding to the clinical annotations of seizures were obtained around the $\alpha$, $\beta$, and  $\gamma$ bands (Fig.~\ref{fig:exp}).
However, the $\gamma$ band showed a relatively high rate of false positives in the post-seizure regions.
This may reflect the muscle activity that occurs after seizures.
The baseline probabilistic models, GHMM and SMM, failed to detect the seizure segments in both examples.
Although MLP and LLR could detect seizure segments, to some extent, the results did not achieve stability and lacked contiguity.
These results indicate that the proposed model can detect seizures more correctly than the baselines by capturing both the stochastic features of EEG and their transitions.

The detection performance of each model is quantified in Table~\ref{tab:results}.
Overall, the proposed model outperformed the baseline models.
High detection ability was achieved in the $\alpha$ and $\beta$ bands for MCC and AUCs, which are comprehensive prediction scores, meaning that the proposed model has a relatively balanced detection ability.
In terms of sensitivity and specificity, the proposed model had a higher sensitivity than the baselines, but lower specificity.
Although there is a trade-off between sensitivity and specificity, a balance between them is generally important.
However, from the viewpoint of the burden on clinical practice, it is necessary to reduce false alarms as much as possible; hence, in the future, the specificity of the proposed model should be improved while maintaining its sensitivity.

In the proposed model, although the $\alpha$ band tended to show the highest detection performance, the $\delta$, $\theta$, and $\beta$ bands scored better in terms of specificity.
In fact, Fig.~\ref{fig:exp} indicated that there were fewer false positives in non-seizure segments in the $\beta$ band.
Fig.~\ref{fig:hist} shows that the best frequency band was different for each subject.
Although a relatively large number of patients showed the best detection performance in the $\alpha$ band, more than half of the patients showed the best performance in the other band, especially the $\theta$ and $\beta$ bands.
The reason for this may be that the frequency characteristics of the EEG activity associated with epileptic seizures differed depending on various factors, such as the age of patients, presence of body movement, and duration of seizures.
Therefore, combining the detection results for multiple frequency bands is expected to improve detection performance.

In this paper, the state-posterior distribution $p(\mathbf{z}|\mathbf{X})$ was estimated for the novel data $\mathbf{X}$ based on the backward-forward algorithm.
Because this algorithm requires future values of data in the estimation of the state at each point, the proposed model is suitable for offline detection.
This is expected to lead to applications such as the automatic detection of the number of seizures and their duration from long-term EEG recordings.
However, in clinical practice, it is also important to detect seizures immediately while monitoring the patient in real time.
This type of application could be achieved by introducing an online prediction algorithm based on the sequential forward calculation.

\section{Conclusion}
In this paper, we proposed a time-series SMM of EEG based on a hidden Markov structure and applied it to seizure detection.
The proposed model can represent the time-series changes of EEG for seizures by switching the distribution of the latent scale parameter based on Markov transitions.
We validated our model on clinical EEG data from 20 focal epileptic patients.
The results demonstrated that the proposed model outperformed several baseline approaches including the GHMM, original SMM, and classical framewise detection methods.
In the future, we will introduce the combination learning of multiple frequency bands and conduct a further evaluation using a greater number of datasets.

\section*{Acknowledgement}
We thank Mr. Ryota Onishi for his help in data cleansing.

\addtolength{\textheight}{-12cm}   


\end{document}